\def\be{\begin{equation}}
\def\te{\end{equation}}
\def\ba{\begin{eqnarray}}

\def\ta{\end{eqnarray}}

\def\be{\begin{equation}}
\def\ee{\end{equation}}
\def\bea{\begin{eqnarray}}

\def\eea{\end{eqnarray}}

\newskip\humongous \humongous=0pt plus 1000pt minus 1000pt

\newif\ifdtup

\def\(#1){(\ref{#1})}

\documentclass[prd,aps,eqsecnum]{revtex4}

\begin{document}

\title{Can Spacetime be a Condensate?}

\author{B. L. Hu}
\email[Electronic address: ]{hub@physics.umd.edu}
\affiliation{Department of Physics, University of Maryland,\\
College Park, Maryland 20742-4111, USA}
\date{May 21, 2005}

\begin{abstract}
We explore further the proposal \cite{GRhydro} that general
relativity is the hydrodynamic limit of some fundamental theories
of the microscopic structure of spacetime and matter, i.e.,
spacetime described by a differentiable manifold is an emergent
entity and the metric or connection forms are collective
variables valid only at the low energy, long wavelength limit of
such micro-theories. In this view it is more relevant to find
ways to deduce the microscopic ingredients of spacetime and
matter from their macroscopic attributes than to find ways to
quantize general relativity because it would only give us the
equivalent of phonon physics, not the equivalents of atoms or
quantum electrodyanmics. It may turn out that spacetime is merely
a representation of collective state of matter in some limiting
regime of interactions, which is the view expressed by Sakharov
\cite{Sak}. In this talk, working within the conceptual framework
of geometro-hydrodynamics, we suggest a new way to look at the
nature of spacetime inspired by Bose-Einstein Condensate (BEC)
physics. We ask the question whether spacetime could be a
condensate, even without the knowledge of what the `atom of
spacetime' is. We begin with a summary of the main themes for
this new interpretation of cosmology and spacetime physics, and
the `bottom-up' approach to quantum gravity. We then describe the
`Bosenova' experiment of controlled collapse of a BEC and our
cosmology-inspired interpretation of its results. We discuss the
meaning of a condensate in different context. We explore how far
this idea can sustain, its advantages and pitfalls, and its
implications on the basic tenets of physics and existing programs
of quantum gravity.
\end{abstract}

\maketitle

\textit{\small - Invited Talk presented at the Peyresq Meetings
2002 - 2003. To appear in Int. J.Theor. Phys. [gr-qc/0503067] }

\section{Introduction}

\subsection{Classical, semiclassical,
stochastic and quantum Gravity}

The theory of general relativity provides an excellent
description of the features of large scale spacetime and its
dynamics. \textbf{Classical gravity} assumes classical matter as
source in the Einstein equation. When quantum fields are included
in the matter source, a \textbf{quantum field theory in curved
spacetimes} (QFTCST) is needed.
\cite{DeW75,BirDav,Fulling,Wald,Mostepanenko,MirVil}. At the
semiclassical level the source in the (semiclassical) Einstein
equation is given by the expectation value of the energy momentum
tensor operator of quantum matter fields with respect to some
quantum state. \textbf{Semiclassical gravity} \cite{scg} refers to
the theory where classical spacetime is driven by quantum fields
as sources, thus it includes the backreaction of quantum fields on
spacetime and self-consistent evolution of quantum field and
spacetime together. Without the requirement of self-consistent
backreaction QFTCST can be viewed as a test field approximation of
semiclassical gravity. \textbf{Stochastic gravity}
\cite{stogra,MarVer,StoGraRev} includes the fluctuations of
quantum field as source described by the Einstein-Langevin
equation \cite{ELE}. Our program on \textbf{quantum gravity} uses
stochastic gravity as a launching platform and kinetic theory
\cite{kinQG} as a program guide.

To anchor our discussions, we summarize here some \textbf{lead
ideas} on the three levels of gravitation theory, shy of quantum
gravity.  We present the main thesis, spell out the major tasks
for each level and new questions which we need to address.

\subsubsection{Cosmology as `Condensed Matter' Physics}

Of importance is not just \textbf{particles and fields} which one
obviously needs for the content of matter which drives the
dynamics of spacetime, but also how they organize and transform
on larger scales. We emphasize the importance of ideas from
\textbf{condensed matter physics}, (in conjunction with quantum
field theory, for treating early universe quantum processes)  in
understanding how spacetime and matter in different forms and
states interplay and evolve. The suggestion of viewing cosmology
in the light of condensed matter physics, in terms of taking the
correct viewpoints to ask the right questions, and approaches to
understand the processes, has been made earlier (e.g.,
\cite{HuHK}). Phase transition processes underlie the foundation
of the inflationary cosmology program. Proposals to study
cosmological defect formation in helium experiments and to view
cosmology as a critical phenomenon were proposed
\cite{ZurekHe,SmoCos}. A recent monograph is devoted to the unity
of forces at work in He$^3$ droplets \cite{Volovik} (see also
\cite{JacVolHe3}).

\subsubsection{General Relativity as Hydrodynamics}

In our view \cite{GRhydro} general relativity is the hydrodynamic
(the low energy, long wavelength) regime of a more fundamental
microscopic theory of spacetime, and the metric and the
connection forms are the collective variables derived from them.
At shorter wavelengths or higher energies, these collective
variables will lose their meaning, much as phonon modes cease to
exist at the atomic scale. This view marks a big divide on the
meaning and practice of quantum gravity. In the traditional view,
quantum gravity means quantizing general relativity, and in
practice, most programs under this banner focus on quantizing the
metric or the connection functions. Even though the stated goals
of finding a microstructure of spacetime is the same, the real
meaning and actual practice between these two views are
fundamentally different. If we view GR as hydrodynamics and the
metric or connection forms as hydrodynamic variables,  quantizing
them will only give us a theory for the quantized modes of
collective excitations, such phonons in a crystal, but not a
theory of atoms or QED. (A similar viewpoint is expressed by
Jacobson \cite{JacEqState} from a different angle. See also
\cite{Pad})

\subsubsection{Stochastic Semiclassical Gravity: Fluctuations and Correlations}

Stochastic semiclassical gravity is a consistent and natural
generalization of semiclassical gravity to include the effects of
quantum fluctuations. The centerpiece of this theory is the
stress-energy bi-tensor and its expectation value known as the
noise kernel. The key point here is the important role played by
noise, fluctuations, dissipation, correlations and quantum
coherence, the central issues focused in and addressed by
mesoscopic physics. This new framework allows one to explore the
quantum statistical properties of spacetime: How fluctuations in
the quantum fields induce metric fluctuations and seed the
structures of the universe, black hole quantum horizon
fluctuations, the backreaction of Hawking radiance in black hole
dynamics, and implications on trans-Planckian physics.  On the
theoretical issues, stochastic gravity is the necessary
foundation to investigate the validity of semiclassical gravity
and the viability of inflationary cosmology based on the
appearance and sustenance of a vacuum energy-dominated phase. It
is also a useful platform supported by well-established low energy
(sub-Planckian) physics to explore the connection with high
energy (Planckian) physics in the realm of quantum gravity.

\subsubsection{`Bottom-up' Approach to Quantum Gravity: Mesoscopic Physics}

As remarked above, we find it more useful to find the
micro-variables than to quantize macroscopic variables. If we
view classical gravity as an effective theory, i.e., the metric
or connection functions as collective variables of some
fundamental particles which make up spacetime  in the large,  and
general relativity as the hydrodynamic limit, we can also ask if
there is a regime like kinetic theory of molecular dynamics or
mesoscopic physics of quantum many body systems intermediate
between quantum micro-dynamics and classical macro-dynamics. This
transition involves both the micro to macro transition and the
quantum to classical transition, two central issues in mesoscopic
physics. We will describe the mesoscopic physics issues here and
the kinetic theory approach to quantum gravity in the next
section.

In \cite{meso} we pointed out that  many issues special to this
intermediate stage, such as the transition from quantum to
classical spacetime via the decoherence of the `density matrix of
the universe', phase transition or cross-over behavior at the
Planck scale, tunneling and particle creation, or growth of
density contrast from vacuum fluctuations, share some basic
concerns of mesoscopic physics in atomic or nuclear condensed
matter or quantum many body systems.   Underlying these issues
are three main factors: quantum coherence, fluctuations and
correlations.  We discuss how a deeper understanding of these
aspects of fields and spacetimes related to the quantum/
classical and the micro /macro interfaces, the discrete /
continuum or the stochastic / deterministic transitions can help
to address some basic problems in gravity, cosmology and black
hole physics such as Planck scale metric fluctuations,
cosmological phase transition and structure formation, and the
black hole entropy, end-state and information paradox.

Mesoscopic physics deals with problems where the characteristic
interaction scales or sample sizes are intermediate between the
microscopic and the macroscopic. The experts refer to a specific
set of problems in condensed matter and atomic / optical physics.
For the present discussion, I will adopt a more general
definition, with `meso' referring to the interface between macro
and micro on the one hand and the interface between classical and
quantum on the other. These two aspects will often bring in the
continuum / discrete and the deterministic / stochastic factors.
These issues concerning the micro / macro interface and the
quantum to classical transition arise in quantum cosmology and
semiclassical gravity in a way categorically similar to the new
problems arising from condensed matter and atomic/optical physics
(and, at a higher energy level, particle/nuclear physics, at the
quark-gluon and nucleon interface). Similarly,  many issues in
gravitation and cosmology are related to the coherence and
correlation properties of quantum systems, and
involve 
stochastic notions, such as noise, fluctuations, dissipation and
diffusion in the treatment of transport, scattering and
propagation processes.

The advantage of making such a comparison between these two
apparently disjoint disciplines is twofold: The  theory of
mesoscopic processes which can be tested in laboratories with the
new nanotechnology can enrich our understanding of the basic
issues common to these disciplines while being extended to the
realm of general relativity and quantum gravity. The formal
techniques developed and applied to problems in quantum field
theory, geometry and topology can be adopted to treat condensed
matter and atomic/optical systems with more rigor, accuracy and
completeness. Many conceptual and technical challenges are posed
by mesoscopic processes in both areas.

\subsection{Geometro-Hydrodynamics: Spacetime as Condensate}

We now present a new idea inspired by the development of
Bose-Einstein Condensate (BEC) physics in recent years. While the
conception of mesoscopic physics and the kinetic theory approach
to quantum gravity bear on the last two themes in the prior
subsection, here we return to the first two themes, dealing with
the hydrodynamic properties of spacetime and their manifestation
in cosmology  through quantum processes involving vacuum
fluctuations. The idea is that maybe spacetime, describable by a
differentiable manifold structure, valid only at the low-energy
long-wavelength limit of some fundamental theory, is a
condensate. We will devote a section examining what a condensate
means, but for now we can use the BEC analog and think of it as a
collective quantum state of many atoms with macroscopic quantum
coherence. {When this thought came to my mind some 5-6 years ago
amidst bursting activities of BEC experiments and theories, I
discarded it immediately for the obvious absurdities indicated
below. After living with this idea for some time they don't seem
as repugnant as before so I dare to share them here in the hope
the audience/reader can throw some much needed light to it}.\\

\subsubsection{Unconventional view 1: All sub-Planckian physics
are low temperature physics}

Atom condensates exist at very low temperatures. It takes novel
ways of cooling the atoms, many decades after the theoretical
predictions, to see a BEC in the laboratories.  It may not be too
outlandish to draw the parallel with spacetime as we see it
today, because the present universe is rather cold (~3K). But we
believe that the physical laws governing today's universe are
valid all the way back to the GUT (grand unification theory) and
the Planck epochs, when the temperatures were not so low any
more. Any normal person would consider the Planck temperature
$T_{Pl}= 10^{32}K$ a bit high. Since the spacetime structure is
supposed to hold (Einstein's theory) for all sub-Planckian eras,
if we consider spacetime as a condensate today, shouldn't it
remain a condensate at this ridiculously high temperature? That
was PUZZLE number 1.

YES is my answer to this question. What human observers consider
as high temperature (such as that when species homo-sapiens will
instantly evaporate) has no effect on the temperature scales
defined by physical processes which in turn are governed by
physical laws. Instead of conceding to a breakdown of the
spacetime condensate at these temperatures, for the sake of
arguments here, one should push this concept to its limit and
come to the conclusion that all known physics today, as long as a
smooth manifold structure remains valid for spacetime, the arena
where all physical processes take place,  are low-temperature
physics. Spacetime condensate exists even at Planckian
temperature $T_{Pl}$, but will cease to exist above the Planck
temperature, according to our current understanding of the
physical laws. In this sense spacetime physics as we know it is
low temperature hydrodynamics, and, in particular, today we are
dealing with ultra-low temperature physics, similar to superfluids
and BECs. \footnote{Other discussions of condensates in gravity
include the proposal of Mazur and Mottola on the existence of
gravitational vacuum condensate stars \cite{MazMot} related to
the earlier work of Chapline and Laughlin et al \cite{Laughlin}
on quantum phase transitions near a black hole horizon. We are
considering the properties of post-Planckian spacetime in general
terms while their considerations predict specific consequences for
unknown and known astrophysical objects. While their general views
in a broader perspective may be considered to be similar to what
is proposed here, we cannot concur with their specific claims.}

The metric or connection functions are hydrodynamic variables,
and most macroscopic gravitational phenomena can be explained as
collective modes and their excitations (of the underlying deeper
micro-theory): from gravitational waves in the weak regime as
perturbations, to black holes in the strong regime, as solitons
(nonperturbative solutions). There may even be analogs of
turbulence effects in geometro-hydrodynamics, when
our observation or numerical techniques are improved.  \\

\subsubsection{Unconventional view 2: Spacetime is, after all, a
quantum entity}

An even more severe difficulty in viewing spacetime as a
condensate is to recognize and identify the quantum features in
spacetime as it exists today, not at the Planck time. The
conventional view holds that spacetime is classical below the
Planck scale, but quantum above. That was the rationale for
seeking a quantum version of  general relativity, beginning with
quantizing the metric function and the connection forms. Our view
is that the universe is fundamentally a quantum phenomena
\footnote{This is still a nascent and very tentative view, I will
explore this idea further in the context of macroscopic quantum
phenomena in \cite{Qupon}}, but at the mean field level the many
body wave functions (of the micro-constituents, or the `atoms' of
spacetime) which we use to describe its large scale behavior
(order parameter field) obey a classical-like equation, similar
to the Gross-Pitaevsky equation in BEC, which has proven to be
surprisingly successful in capturing the large scale collective
dynamics of BEC \cite{PethickSmith}, until quantum fluctuations
and strong correlation effects enter into the picture
\cite{ReyHFB}.

Could it be that the Einstein equations depict the collective
behavior of the spacetime quantum fluid  on the same footing as a
Gross-Pitaevsky equation for BEC? The deeper layer of structure
is ostensibly quantum, it is only at the mean field level that
the many-body wave function is amenable to a classical
description. We have seen many examples in quantum mechanics
where this holds,  In truth, for any quantum system which has
bilinear coupling with its environment or is itself Gaussian
exact (or if one is satisfied with a Gaussian approximation
description) the equations of motion for the expectation values of
the quantum observables have the same form as its classical
counterpart. The Ehrenfest theorem interwoven between the quantum
and the classical is one common example.

The obvious challenge is, if the universe is intrisically quantum
and coherent, where can one expect to see the quantum coherence
phenomena of spacetime? Here again we look to analogs in BEC
dynamics for inspiration, and there are a few useful ones, such as
particle production in the collapse of a BEC,  which we will
describe in a later section. One obvious phenomenon staring at our
face is the vacuum energy of the spacetime condensate, because if
spacetime is a quantum entity, vacuum energy density exists
unabated for our present day late universe,  whereas its origin is
somewhat mysterious for a classical spacetime in the conventional
view. We'd like to explore the implications of this view on the
cosmological constant and coincidence problems later.
\footnote{Volovik has proposed some solutions to these problems.
While we agree with his general attitude we reserve our judgement
on the particulars in the theories and models he proposed. The
issues are subtle and complex.  See also Padmanabhan's views in
\cite{Pad1}.}

In the next section we summarize the `kinetic theory approach to
quantum gravity, as it is one way to connect the (macro)
hydrodynamics  to the (micro) molecular dynamics. \footnote{For
the last two decades working on these ideas I have been inspired
by work of \cite{molhydro} on the relation between hydrodynamics
and molecular dynamics and \cite{HydroFluc,KacLogan,Lax} on
hydrodynamic fluctuations. To see how many-atom correlations
interplay with hydrodynamical features of BEC via a kinetic
theory description, see, e,g, \cite{ReyHFB,ReyKin}}. In Sec. IV
we address the main issues associated with the spacetime
condensate viewpoint, taking on its meaning and discussing its
implications on the basic tenets of physics and existing programs
of quantum gravity.



\section{From Stochastic to Quantum Gravity via Metric Correlation Hierarchy}

In this section we summarize the main points in the kinetic theory
approach to quantum gravity \cite{kinQG}.  Again, by quantum
gravity we mean a theory of the microscopic structure of
spacetime, not necessarily a theory obtained by quantizing
general relativity. The key ideas utilized to construct this
proposal are the  correlation hierarchy \cite{Balescu,CH88},
decoherence of correlation history \cite{dch}, correlation noise
\cite{cddn} and stochastic Boltzmann equation \cite{stobol}. In
statistical physics it is well-known that intermediate regimes
exist between the long-wavelength hydrodynamics limit and the
microdynamics \footnote{They are usually lumped together and
called the kinetic regime, but I think there must be distinct
kinetic collective variables depicting recognizable metastable
intermediate structures in this vast interim regime (see
\cite{Spohn})}. The central task for us is the retrieval or
reconstruction of quantum coherence in the gravity sector. We do
this through fluctuations and correlations, starting from the
matter sector described by quantum fields, and connecting to the
gravity sector by the Einstein equations, at the hydrodynamic
level, and its higher order hierarchical generalizations, at the
kinetic theory level. The pathway from stochastic to quantum
gravity \footnote{For discussions on this more general issue,
see, e.g., \cite{Accardi,CharisQSto,CRVphysica}} in the kinetic
theory approach is via the correlation hierarchy of noise and
induced metric fluctuations. Readers who are familiar with this
can skip to the next Section.

We see that stochastic semiclassical gravity provides a relation
between noise in quantum fields and metric fluctuations. While the
semiclassical regime describes the effect of a quantum matter
field only through its mean value (e.g., vacuum expectation
value), the stochastic regime includes the effect of fluctuations
and correlations. We believe that precious new information
resides in the two-point functions and higher order correlation
functions of the stress energy tensor which may reflect the finer
structure of spacetime at a scale when information provided by
its mean value as source (semiclassical gravity) is no longer
adequate.

Our strategy is to look closely into the quantum and statistical
mechanical features of the matter field in deepening levels and
see what this implies on the spacetime structure at the
corresponding levels. (This is different from the induced gravity
program \cite{Sak} although the spirit is similar). Thus we work
with both the micro structure of matter described by quantum
field theory of matter and the macro structure of spacetime
described by hydrodynamics.  We rely on higher order correlations
in moving beyond the semiclassical gravity stage. The procedures
in this approach involve the deduction of the correlations of
metric fluctuations from correlation noise in the matter field,
identifying distinct collective variables depicting recognizable
metastable structures in the kinetic and hydrodynamic regimes of
quantum matter fields and finding out the corresponding structure
and behavior in their  spacetime counterparts.

This will give us a hierarchy of generalized stochastic
equations, the Boltzmann-Einstein hierarchy of quantum gravity,
for each level of spacetime structure, from the macroscopic
(general relativity) through the mesoscopic (stochastic gravity)
to the microscopic (quantum gravity). The linkage at the lowest
level is provided by the Einstein equation. Stochastic gravity
entails all the higher rungs between semiclassical and quantum
gravity, much like the BBGKY \cite{Balescu} or the Dyson-Schwinger
hierarchy \cite{stobol} representing kinetic theory of matter
fields.


\subsection{Noise and Fluctuations as Measures of Correlations and Coherence}

In \cite{stogra} a simple example was given to illustrate the
relation of the stochastic regime compared to the semiclassical
and the quantum. We see that (at least for linear gravitational
perturbations) the stochastic equations contain the same
information as in quantum gravity, with the quantum average
replaced by the noise average. (See also
\cite{CharisQSto,CRVphysica}) The difference is that for
stochastic gravity the average of the energy momentum tensor is
taken with respect only to the matter field, but not the graviton
field. The important improvement over semiclassical gravity is
that it now carries information on the correlation (and the
related phase information) of the energy momentum tensor of the
fields and its induced metric fluctuations which is absent in
semiclassical gravity. (The relation between {\it fluctuations}
and {\it correlations} is a variant form of the
fluctuation-dissipation relation.) The correlation in quantum
field and geometry fully present in quantum gravity yet
completely absent in semiclassical gravity, is partially captured
in stochastic gravity. It is in this sense that a stochastic
gravity gives a much improved description and is closer to the
quantum than the semiclassical.

Noise or fluctuations  holds the key to probing the quantum
nature of spacetime in this vein. The background geometry is
affected by the correlations of the quantum fields through the
noise term in the Einstein-Langevin equation, manifesting as
induced metric fluctuations. The Einstein-Langevin equation in
the form written down in \cite{ELE} contains only the lowest
order term, i.e., the 2 point function of the energy momentum
tensor (which contains the 4th order correlation of the quantum
field, or gravitons, when they are considered as matter source).
\footnote{Although the Feynman- Vernon scheme can only accomodate
Gaussian noise of the matter fields and takes a simple form for
linear coupling to the background spacetime, the notion of noise
can be made more general and precise. For an example of a more
complex noise associated with more involved backreactions arising
from strong or nonlocal couplings, see \cite{JH1}} Noise in a
broader sense embodies the contributions of the higher
correlation functions in the quantum field. One could deduce
generalized Einstein-Langevin equations containing more complex
forms of noise, which fall under the same stochastic gravity
programatic scheme. Progress is made on how to characterize the
higher order correlation functions of an interacting quantum field
systematically from the Schwinger-Dyson equations in terms of
`correlation noises' \cite{cddn,stobol}, similar to the classical
BBGKY hierarchy.

One can generalizing this scheme to the gravity-matter system,
viewed as a system of strongly interacting fields, towards a
description of the micro-structure of spacetime. Starting with
stochastic gravity we can get a handle on the correlations of the
underlying field of spacetime by examining (observationally if
possible, e.g., effects of induced spacetime fluctuations) the
hierarchy of equations, of which the Einstein-Langevin equation
given in \cite{ELE} is at the lowest order, i.e., the relation of
the mean field to the two point function, and the two point
function to the four (variance in the energy momentum tensor),
and so on. One can in principle move up in this hierarchy to probe
the dynamics of the higher correlations of spacetime
substructure. This is the basis for a correlation dynamics
/stochastic semiclassical approach to quantum gravity
\cite{stogra}.

\subsection{Quantum Coherence in the Gravity Sector Obtained
from Correlations of Induced Metric Fluctuations}

Noise carries information about the correlations of the quantum
field. One can further link {\it correlation} in quantum fields
to {\it coherence} in quantum gravity. This linkage is ensured in
principle, by virtue of the fact that at the quantum gravity level
a complete quantum description should be given by a coherent wave
function of the combined matter and gravity sectors. This linkage
is operationally viable because of the self-consistency
requirement (full backreaction is included) in the Einstein
(classical level), the semiclassical Einstein (semiclassical
level) and the Einstein-Langevin equations (the stochastic level)
which relate the matter and spacetime sectors at the respective
levels. Semiclassical gravity does not contain any information
about the quantum coherence in the gravity sector. Stochastic
gravity improves on the semiclassical in that it preserves
partial information related to the quantum coherence in the
gravity sector, by including the correlations in the matter field
which contains quantum coherence information.

Since the degree of coherence can be measured in terms of
correlations, our strategy towards quantum gravity in the
stochastic gravity program is to unravel the higher correlations
of the matter field, go up the hierarchy starting with the
variance of the stress energy tensor,  and through its linkage
with gravity (the lowest rung provided by the Einstein equation),
retrieve whatever quantum attributes (partial coherence) of
gravity left over from supra-Planckian high energy behavior. Thus
in this approach, focusing on the noise kernel and the stress
energy tensor two point function is our first step beyond the
mean field theory (semiclassical gravity) towards probing the
full theory of quantum gravity.

We have only addressed the correlation aspect; there is also the
quantum to classical aspect. One way to address this issue is by
the decoherence of correlation histories scheme proposed in
\cite{dch}, another is by the large N approximation. \footnote{In
\cite{stogra} I also brought up the relevance of the large N
expansion in gravity for comparison. There exists a relation
between correlation order and the loop order \cite{cddn}. One can
also relate it to the order in large N expansion (see, e.g.,
\cite{AarBer}). It has been shown that the leading order 1/N
expansion for an N-component quantum field yields the equivalent
of semiclassical gravity \cite{HarHor}. The leading order 1/N
approximation yields mean field dynamics of the Vlasov type which
shows Landau damping which is intrisically different from the
Boltzmann dissipation. In contrast the equation obtained from the
nPI (with slaving) contains dissipation and fluctuations
manifestly. It would be of interest to think about the relation
between semiclassical and quantum in the light of the higher 1/N
expansions \cite{Tomb}, which is quite different from the
scenario associated with the correlation hierarchy. The next to
leading order calculation has recently been performed by Roura
and Verdaguer \cite{RVlargeN}.}

\subsection{Spacetime as an Emergent Collective State of Strongly Correlated Systems}

At this point it is perhaps useful to revisit an earlier theme we
presented in the beginning, i.e., Stochastic semiclassical
gravity as mesoscopic physics.

Viewing the issues of correlations and quantum coherence in the
light of mesoscopic physics we see that what appears on the right
hand side of the Einstein-Langevin equation, the stress-energy two
point function, is analogous to conductance of electron transport
which is given by the current-current two point function. What
this means is that we are really calculating the transport
functions of the matter particles as depicted here by the quantum
fields. Following Einstein's observation that spacetime dynamics
is determined by (while also dictates) the matter (energy
density), we expect that the transport function represented by
the current correlation in the fluctuations of the matter energy
density would also have a geometric counterpart and equal
significance at a higher energy than the semiclassical gravity
scale. This is consistent with general relativity as
hydrodynamics: conductivity, viscosity and other transport
functions are hydrodynamic quantities. Here we are after the
transport functions associated with the dynamics of spacetime
structures. The Einstein tensor correlation function calculated
by Martin and Verdaguer \cite{MarVer} is one such example.
Another example is in the work of Shiokawa on mesoscopic metric
fluctuations \cite{Shiok}.

For many practical purposes we don't need to know the details of
the fundamental constituents or their interactions to establish
an adequate depiction of the low or medium energy physics, but
can model them with semi-phenomenological concepts. When the
interaction among the constituents gets stronger, or the probing
scale gets shorter, effects associated with the higher
correlation functions of the system begin to show up. Studies in
strongly correlated systems are revealing in these regards.
Thus, viewed in the light of mesoscopic physics, with stochastic
gravity as a stepping stone, we can begin to probe into the
higher correlations of quantum matter and with them the
associated excitations of the collective modes in
geometro-hydrodynamics, the kinetic theory of spacetime
meso-dynamics and eventually quantum gravity -- the theory of
spacetime micro-dynamics.

In seeking a clue to the micro theory of spacetime from
macroscopic constructs, we have focused  here on the kinetic /
hydrodynamic theory and noise / fluctuations aspects. Another
equally important factor is topology. Topological features can
have a better chance to survive the coarse-graining or effective /
emergent processes to the macro world and can be a powerful key
to unravel the microscopic mysteries. This aspect is left for
future discussions.


\section{What can we learn about quantum gravity from BEC}

In the Introduction we have stated the main theme of considering
spacetime as a condensate, and mentioned several puzzles and
challenges such a view evokes. We shall elaborate on those points
in this section. But before doing so, we want to augment our
physical intuition with a description of an analogy between
phenomena observed in BEC collapse experiments \cite{JILA01b},
and quantum field processes in the early universe. This
observation was made in a recent work
\cite{CHbosenova} 
\footnote{We wish to mention other black hole \cite{Garay} and
cosmological \cite{FF03} analog studies of BEC.} The main
features are described below.


\subsection{Vacuum Cosmological Processes found in Controlled BEC collapse}

We show that in the collapse of a Bose-Einstein condensate (BEC)
certain processes involved and mechanisms at work share a common
origin with corresponding quantum field processes in the early
universe such as particle creation, structure formation and
spinodal instability. Phenomena associated with the controlled
BEC collapse observed in the experiment of Donley et al
\cite{JILA01b} (they call it `Bose-Nova', see also \cite{CVK03})
such as the appearance of bursts and jets can be explained as a
consequence of the squeezing and amplification of quantum
fluctuations above the condensate by the dynamics of the
condensate.

The collapsing BEC is the time-reverse scenario of an expanding
universe and the condensate plays a similar role as the vacuum in
quantum field theory in curved spacetime.  One can understand the
production of atoms in the form of jets and bursts as the result
of parametric amplification of vacuum fluctuations by the
condensate dynamics. This is the same mechanism as cosmological
particle creation from the vacuum, which is believed to be
copious near the Planck time. Some basic ideas common to
cosmological theories like ``modes freeze when they grow outside
of the horizon" can be used to explain the special behavior of
jets and bursts ejected from the collapsing BEC. Finally the
waiting time before a BEC starts to collapse obey a scaling rule
which can be derived from simple principles of spinodal
instability in critical phenomena.

Using the physical insight gained in depicting these cosmological
processes, our analysis of the changing amplitude and particle
contents of quantum excitations in these BEC dynamics provides
excellent quantitative fits with the experimental data on the
scaling behavior of the collapse time and the amount of particles
emitted in the jets. Because of the coherence properties of BEC
and the high degree of control and measurement precision in
atomic and optical systems, we see great potential in the design
of tabletop experiments for testing out general ideas and
specific (quantum field) processes in the early universe, thus
opening up the possibility for implementing `laboratory
cosmology'.


\subsection{What is a condensate?}

We have mentioned BEC as an example of a condensate. The spectrum
is much broader. We now give a more systematic description of it.
We will see the differences between photons and gravitons versus
bosonic atoms in BEC; particles versus quasiparticles and
collective excitations.\\

\paragraph{Condensate as a "macroscopically populated" coherent
state} Under this category are (i) "classical" electromagnetic
wave, which can be thought of as a photon condensate; (ii)
"classical" elastic wave as a phonon condensate; (iii) "classical"
gravitational wave, a graviton condensate. Note that they are all
coherent.

\paragraph{Condensate as a non-trivial equilibrium phase at T=0} This
is the case for (i) BEC, as far as the bosonic atoms are
concerned and (ii) BCS state, as far as the fermionic atoms are
concerned, but NOT for gravitons. In fact, the situation for
gravitons is similar to (i)
photons (ii) phonons (iii) the quasiparticles in a BEC.\\

We can now better define the meaning of a condensate in the
following questions:

\textit{c.} ``Can spacetime be considered a condensate from some
microscopic (more fundamental) substructure, so that the metric
and its perturbations correspond to collective variables and
collective excitations?'' This is similar to vibrational modes
(phonons) in a lattice of atoms, vibrational and rotational modes
of a nucleus (nuclear collective model), BEC quasiparticles or
He$^4$ superfluid dynamics (This is the picture behind
\cite{GRhydro})

\textit{d.} One can also think of the condensate as a nontrivial
quantum state in terms of the microscopic constituents, such as
in the string theory picture (see discussions below). This is
similar to the ground states in (ii) BEC (iii) BCS (both
involving non-trivial Bogoliubov transformations), but different
from (i) phonon vacuum in a lattice (just normal modes). We need
a microscopic theory to distinguish these two cases, or more
information about the structures arising from graviton-graviton nonlinear interactions.\\

Finally we can ask,

\begin{enumerate}

\item ``Is there any way to rule out the possibility that the
graviton vacuum (for different background geometries) is a
condensate in the sense of either case c) or d) above? We cannot
think of a way to do so yet. One should think harder to either
substantiate or falsify this view.

\item Are there any hints suggesting that this is a possibility? --
Maybe. Examples are:

(i) Trans-Planckian modes in black holes horizons \cite{transPl}

(ii) Black hole atom \cite{BHatom}, Black hole quasi-normal modes
\cite{BHqnm} and Black hole event horizon fluctuations
\cite{BHhorfluc}

(iii) Cosmological constant problem.

This viewpoint may provide a more natural explanation of the dark
energy mystery: Why is the cosmological constant so low (compared
to natural particle physics energy scale) today, and so close to
the matter energy density?

\end{enumerate}

Using these finer distinctions it is worthy to explore the
implications and contradictions from this viewpoint of a
spacetime condensate.



\section{Spacetime condensate viewpoint: Implications on basic principles}


\subsection{Comparison with the proposals of Volovik}

As mentioned in the Introduction, the body of work by Volovik can
probably be perceived as closest to our view here. For this
reason it is perhaps useful to delineate the similarities and
differences. Put broadly, we would say that the general
philosophy and perspective are similar. (So is with Jacobson's
\cite{JacEqState}), but differences exist in the working
principles or the choice of models. While we admire the boldness
in Volovik's proposals, we would exercise caution in making
certain sweeping claims. Nevertheless, the points of agreements
are more basic and concordent:

1. Low energy properties of different vacua are robust:  magnets,
superfluids, crystals, liquid crystals, superconductors. They do
not depend much on the details of micro-structure, i.e., atoms.

2. Microphysics only provides the constants of macrophysics:
speed of sound, superfluid density, modulus elasticity, magnetic
susceptibility. In our view, these are all derived properties of
an emergent structure. They are not fundamental in the sense that
there are microscopic structures beneath.

3.  Principal role played by symmetry and topology

4.  Different universality classes dictate different behaviors.
One could in principle deduce the properties of vacuum energy
e.g., it is zero and non-gravitating \cite{Weinberg98}

\subsection{Implications on Quantum Mechanics and General Relativity}

The attitudes towards these two pillars of modern physics, quantum
mechanics (QM) and general relativity (GR), are as varied as
there are original thinkers. As a useful contrast, I mention two
views very different from this one presented here, 1) The first
group, represented by Penrose, is willing to give up quantum
mechanics but holds on to GR; 2) Our view here regarding
spacetime as an emergent entity in the low energy limit leads us
to give up on GR beyond the Planck scale when the deeper level of
microstructure of spacetime and matter reveals itself; and 3) the
third group, spearheaded by 'tHooft \cite{tHooftQM}, views
quantum mechanics not as a fundamental theory but as a set of
bookkeeping rules.

1. Those in the first group regarding general relativity as the
deeper theory -- more foundational and elemental -- are ready to
give up on quantum mechanics. In particular, Penrose consigns
gravity the role of facilitating the decoherence of macroscopic
quantum phenomena which shapes the classical world.

2. In this view, GR is only an effective theory valid in the
macroscopic limit. Lorentz invariance and gauge principles are
emergent symmetries. Quantum mechanics governs the micro
structures (atoms, strings) and, as expressed in this talk, may
even govern the macro structures, as collective phenomena.
(quasi-particles, condensates)

3. According to 'tHooft quantum mechanics should be viewed as
dissipative classical dynamics. One apparent difficulty of this
view is in the interpretation of dissipative processes (and the
arrow of time issue) in the context of time-reversal invariant
laws in relation to the basic tenets of statistical mechanics.
One very interesting thought (to this author at least) is that
quantum mechanics is a set of rules which provides an efficient
bookkeeping scheme in our perception of the classical world.
\footnote{This position on the classical-quantum dichotomy is
almost the reverse of the decoherence viewpoint (shared by this
view). The adherents of quantum mechanics would consider the
classical world as an emergent entity  from the decoherence
(environment-induced or consistent-history) of a microscopic
world governed by quantum dynamics.} This is a probe into the
nature of quantum mechanics. If true his viewpoint would demote
the role of quantum mechanics from a fundamental theory of nature
to a scheme, a clever scheme nonetheless, of bookkeeping. By no
means does it imply quantum mechanics is `wrong' -- because it
has proven to work in the physical world.

\subsection{What is the Atom of Spacetime? Implications in relation to String Theory and Loop Quantum Gravity}
Let us now turn to the tough question:
\subsubsection{``What is the Atom of Spacetime?" }

How do we see or find them? In BEC the answer is obvious. BEC is
made from atoms so it is not difficult or surprising to find
atoms originating from, and interacting with, the BEC. Indeed, in
the BEC experiments, when the vacuum (condensate) is squeezed by
a controlled collapse, atoms appear in  bursts or jets (see,
e.g.,\cite{JILA01b}). But we should be mindful that not just
atoms are produced: At a different magnetic field range,
molecules are produced, as evidenced from Ramsey fringes of
molecule- condensate resonances. At higher energies one can
produce other energetic particles. Going beyond the confines of
atomic physics, at nuclear energies one can think of quark gluon
plasma and their condensates \cite{QGcond}. At SUSY scales, one
can think of Higgs condensates.  String condensates, e.g., of
ghosts \cite{GhostCondensate}, if they exist, will also count as
forms of matter structure, albeit at a much deeper level. Thus
there could be as many condensates as there are different levels
of matter or particle structure they are made of.

At today's low energy the information of the detailed composition
is grossly coarse-grained. Only the stress energy tensor of matter
is needed to determine the large scale curvature of spacetime.
Thus one cannot attribute a unique type of condensate which makes
up the spacetime macro-structure as we see it today. Condensates
at all levels of matter structure can contribute, probably with a
weighing factor depending on their spectral distribution which
varies with energy.

Spacetime's geometric description is possible only in the low
energy long wavelength limit.  Beyond the hydrodynamic regime
there may exist as many different mesoscopic regimes for
spacetime structures as there are the corresponding condensates.
None of the low energy or ultra-low temperature condensates
could, by themselves, reveal the atomic structure of spacetime.
But maybe in the squeezing the vacuum (as during rapid expansions
of the early universe in analogy to the Bosenova experiment) or
`tearing up' the spacetime manifold (as in crossing shock waves
or in black hole collisions)  a deeper layer of structure may
reveal. This is one of the motivations behind exploring possible
kinetic theory regimes between the hydrodynamics (General
Relativity) of spacetime structure and the molecular dynamics of
quantum gravity. \footnote{Castro \cite{Castro} claims that he
knows what the 'Atoms or Quanta' of Spacetime are: ``bubbles "
(or p-loops ) of hypervolumes.}

\subsubsection{Implications for String Theory: Can spacetime be derived from strings}

How does the basic premises of string theory fit into this
picture? It does, in the sense that general relativity has been
shown to be the low energy limit of string theories \cite{GSW}.
Whether spacetime is the hydrodynamic limit of string theory has
yet to be shown, but it is believed to be plausible
\cite{Herzog}. This is not so trivial an issue as it may seem,
because so far most discussions of string theory  still assume a
background spacetime where the strings propagate and interact.
(String cosmology certainly makes such an assumption,  when the
line element of a FRW or de Sitter universe is written down.) The
real challenge is for the interacting strings to produce a
spacetime, or at least to see spacetime emerge in some parameter
range of their interactions. The advent of D-branes
\cite{Polchinsky} and duality relations greatly simplify and
organize the structure of string theory, with five interconnected
types, all manifestations of the one M-theory. Discovery of the
AdS/CFT correspondence \cite{Maldacena} changed the perspective
and emphasis significantly. Now one can say that what happens at
the gauge theory (CFT) sector has an exact correspondence in the
spacetime (AdS) sector. In fact it is very interesting because
one can find out the strong coupling regime of gravity from the
weak coupling regime of gauge theories. But can one say that one
sees the emergence of spacetime? Perhaps. Perhaps not, because
the two different regimes are for two different entities
(spacetime and gauge theory). The correspondence provides
interesting and important connections of \textbf{known} physics,
such as QCD and GR (e.g., deconfinement transition in QCD linked
to Hawking-Page transition of black hole nucleation from thermal
AdS space). New physics operative at trans-Planckian scales is
still elusive. The belief is that if we know how to find
solutions to more types of string theory or if we can formulate
string field theory, new physics will appear. Or, perhaps there
is no need or no room for trans-Planckian physics because of the
IR/UV duality. I think it is fair to say that the structural
relation of spacetime and strings remains an open question and a
weighty issue.

\subsubsection{Implications for Loop Quantum Gravity: Is spin connection a fundamental or collective variable?}

The discovery of the Ashtekar variables \cite{Ash} was viewed as
an important step for solving the Einstein constraint equation in
quantum general relativity. Indeed the focus is on quantizing the
spin connection. Another important step in this program which
lends its current name is in recognizing the significance of
Wilson loops \cite{RovSmo} in the loop (Faraday) formulation of
gauge theories. For recent developments, see \cite{LoopQG}.

Considering the quantization of gauge theories in relation to our
view of GR as hydrodynamics, two questions naturally arise, one is
for the loop gravity program:  1) Is the gauge connection a
fundamental or collective variable? This has important
implications in the true value of such a program in quantum
gravity.
The other question is for this geometro-hydrodynamics program.
Since gauge theories share a similar structure as general
relativity, if one regards the connection form in GR as a
collective variable,  how should one view it in gauge theories,
such as the electromagnetic potential? 2) Wouldn't one then
regard all gauge bosons as collective variables and gauge
symmetries as emergent properties particular to these variables?
This is a daring challenge this program raises. For adherents of
this program the logical answer to the second question should be
YES. Then one would be faced with the difficult task of finding
composite or emergent properties for what we would usually regard
as ostensibly elementary particles, like photons. In this light,
the recent proposal of string nets and quantum order by Wen is of
unusual fundamental significance \cite{Wen}. According to his
theory, the collective excitations in string-net condensed phase
can behave just like light and electrons in our vacuum. This
suggests that light and electrons as well as other elementary
particles may originate from string-net condensation in our
vacuum. This is a logical requisite of the idea that all gauge
bosons (expressed in terms of connection forms) are collective
entities. If string-net condensates are found, the discovery will
lend strong support to the spacetime condensate idea, which will
have far-reaching consequences in theoretical physics across the board. \\

{\bf Acknowledgement} I wish to thank Esteban Calzetta for very
fruitful collaborations on the BEC collapse problem and its
cosmological analog, and Albert Roura for very helpful
discussions on the spacetime condensate idea.  I thank Emil
Mottola for reminding me of his work with Pawel Mazur, and
Padmanabhan for his after the first version of this essay
appeared in the arXiv. I thank Pawel Mazur for communicating to
me his interesting ideas.  I enjoyed the hospitality of Edgard
Gunzig at the Peyresq meetings where some key ideas of this paper
was first presented \footnote{The latest development of these
ideas can be found in my talk at QUPON \cite{Qupon}.}. This
research is supported in part by NSF grant PHY03-00710.


\end{document}